\documentclass[twocolumn,showpacs,rotate,aps,prb,floatfix]{revtex4}
\usepackage{epsfig,color}

\def\beq{\begin{equation}}

\def\eeq{\end{equation}}
\def\bea{\begin{eqnarray}}
\def\eea{\end{eqnarray}}

\newcommand{\R}{{\bf r}}

\def\k{\mathbf{k}}
\def\r{\mathbf{r}}
\def\R{\mathbf{R}}

\def\la{\langle}
\def\ra{\rangle}
\def\pp{\prime \prime}
\def\vr{\mathrm{r}}

\input{tcilatex}

\begin{document}

\title{One-Electron Physics of the Actinides}

\author{A.~Toropova}%
\affiliation{Center for  Materials Theory, Department of Physics
and Astronomy, Rutgers University, Piscataway, NJ 08854}

\author{C.~A.~Marianetti}%
\affiliation{Center for  Materials Theory, Department of Physics
and Astronomy, Rutgers University, Piscataway, NJ 08854}

\author{K.~Haule}%
\affiliation{Center for  Materials Theory, Department of Physics
and Astronomy, Rutgers University, Piscataway, NJ 08854}

\author{G.~Kotliar}%
\affiliation{Center for  Materials Theory, Department of Physics
and Astronomy, Rutgers University, Piscataway, NJ 08854}

\date{\today}

\begin{abstract}
We present a detailed analysis of the  one-electron physics of the
actinides. Various LMTO basis sets are analyzed in order to
determine a robust bare Hamiltonian for the actinides. The
hybridization between $f$- an $spd$- states is compared with the
$f-f$ hopping in order to understand the Anderson-like and
Hubbard-like contributions to itineracy in the actinides. We show
that both contributions decrease strongly as one move from the
light actinides to the heavy actinides, while the Anderson-like contribution dominates in all cases. A real-space analysis of
the band structure shows that nearest-neighbor hopping dominates
the physics in these materials. Finally, we discuss the implications of
our results to the delocalization transition as  function of atomic number across the
actinide series.
\end{abstract}

 \maketitle

\section{Introduction}

\subsection{Background of the actinides}

It is well accepted\cite{Pu_Futures} that the actinides are
divided into two groups based on the behavior of the
$f$-electrons. The lighter actinides (Th to Pu) have smaller
atomic volumes, low-symmetry crystal structures and itinerant $5f$
states that participate in metallic bonding\cite{Soderlind95}.
Alternatively, the heavy actinides (Am to Es) have larger atomic
volumes, high-symmetry crystal structures, and relatively
localized $f$-electrons. Applying pressure to the heavy actinides
results in a series of crystallographic phase transitions, and the
respective phases often have significantly different volumes
\cite{Heathman2005,Lindbaum2001}. Transitions of this sort are
often referred to as "volume collapse" transitions.  Given that
the application of ample pressure to any system of localized
electrons will eventually cause a delocalization transition,
understanding what role the electronic delocalization transition
may play in the volume collapse transition has been and continues
to be an active area of study \cite{Savrasov2001,Soderlind2004}.

Plutonium is considered to be the dividing line of actinide series,
with the $\alpha$- and $\delta$- phases associated with light and
heavy behavior, respectively. This dual nature of Pu, along with
an enormous $25\%$ volume collapse for the  $\delta \rightarrow
\alpha$ transition, has made Pu the most interesting element among the
$5f$ compounds for basic theoretical research over the past 50
years \cite{Freeman84,Lashley2005}.

The actinides are among the most complicated classes of materials
in terms of understanding electronic correlations given the
presence of $s$, $p$, $d$, and $f$ electrons near the Fermi
surface and the unusual behavior observed in experiment. Broad
discussion in the literature was devoted to the following
topics: abrupt change in volume and bulk
modulus\cite{Skriver78}; unique crystal
structures\cite{Soderlind95}; partial localization of
$f$-electrons\cite{Eriksson99}, Mott transition
\cite{Johansson74,Johansson75}; paramagnetism in light actinides
and formation of magnetic moments in heavier actinides (starting
from Cm)\cite{Lashley2005}. For this purpose numerous {\it Ab
Initio} electronic structure calculations have been performed for
the actinides. The techniques such as LDA \cite{Ek93,Jones2000},
GGA\cite{Soderlind97}, SIC-LSD\cite{Svane2007},
LDA+U \cite{Savrasov2000,Shorikov05,Shick2005} and DMFT-based
approaches \cite{Zhu2007,Shim2006,Pourovskii2007} have been
implemented\cite{Anisimov2006}.

\subsection{Actinide Hamiltonian}
The model Hamiltonian for the actinides can be written as: \bea
\label{model_Hamiltonian} H =  \sum_{ ij a \beta} V_{ij a \beta}(
c^{\dagger}_{a i} f_{\beta j} + c.c.) + \sum_{ ij \alpha \beta}
t^f_{ij\alpha\beta} f^{\dagger}_{\alpha i} f_{\beta j}
\\ \nonumber
+\sum_{\k ab}t^{spd}_{ab}(\k)c^{\dagger}_{\k a}c_{\k b}
+\sum_{i\alpha\beta\gamma\delta}U_{\alpha\beta\gamma\delta}
f^{\dagger}_{\alpha i}f^{\dagger}_{\beta i}f_{\gamma i}f_{\delta
i} \eea where $f_{\alpha i}$ is the annihilation operator for
$5f$-electron in state $\alpha = |j, j_z \ra $ at site $i$, and
$c_{a i}$ is annihilation operator for conduction electrons in the
state $a = |n, j^{\prime}, j^{\prime}_s \ra $. This model can be
understood as a periodic Anderson model in which additional direct
hopping is allowed between the correlated states. Equivalently,
this model can be thought of as a Hubbard model with additional
uncorrelated states that hybridize with the correlated states.
Therefore, the model for the actinides contains the physics of
\emph{both} the periodic Anderson model and the Hubbard model. In
the Hubbard model $t^f$ competes with $U$ to determine the degree
of localization of the electrons, while in the periodic Anderson
model $V$ competes with $U$. In the model of the actinides $t^f$
and $V$ cooperatively compete with $U$, and the relative
magnitudes of $t^f$ and $V$ will determine the degree of
Hubbard-like and Anderson-like contributions to the itineracy of
the $f$-electrons. The main focus of this study is to determine
the relative importance of $t^f$ and $V$ across the actinide
series. This is a  first step towards a detailed  understanding of
the quantitative aspects of the localization-delocalization
transition in the actinide series.

Hamiltonians of the form described in
Eq.~(\ref{model_Hamiltonian}) containing heavy and light electrons
have appeared in various contexts in condensed matter physics. To
deal with this complexity, this model is often reduced to a
simpler model by eliminating the light electrons  (i.e. $spd$
states) to obtain an effective Hubbard model having only  the
heavy (i.e. $f$) states. The reduced  model  only describes the
bands within a narrow energy window around the Fermi energy (i.e.
$\sim$1eV). The new renormalized $f$ hoppings have contributions
from both the original $f$ hopping in addition to the $spd$
hopping. Additionally, many new interactions terms are generated,
but these are usually ignored and only the on-site coulomb
repulsion is retained as in the original hamiltonian. This
procedure of one-electron "downfolding" has been used extensively
to study electronic correlations in the transition-metal oxides,
and there has been success in describing the photoemission spectra
in this manner. In the context of the actinides, it is not clear
that this approach is justified. Perhaps the most important issue
is the nature of the localization transition in the actinide
series.  In the Hubbard model, when $U$ is sufficiently large the
effective $f$ states will be localized and the system will be
insulating. In the actinide model, when $U$ is sufficiently large
the $f$ states will be localized, but the system will not
necessarily be an insulator.  It is possible that the $spd$ states
may still form a Fermi surface and give rise to a metallic state.
Therefore, the actinide model and the effective Hubbard model
differ even at a qualitative level in certain regimes. Some
aspects of localization-delocalization in the actinide model and
the Hubbard model treated by DMFT are very similar at intermediate
temperatures \cite{Held2000} (for example they both exhibit a line
of first order phase transitions ending at a second order point),
but there are significant differences at very low temperatures
when hybridization becomes a relevant perturbation suppressing
Mott transition \cite{Medici2005}. Furthermore, the behavior at
large $U$ and  high temperatures should be quite different in the
two models, due to the presence of the broad metallic $spd$ bands
in the actinide model. Due to these considerations, we pursue the
actinide model which should be a more accurate representations of
the actinides given that the Hubbard model is an approximate
reduction of the actinide model.

In general, the parameters $V$ and $t^{f}$ depend on the choice of
basis set and therefore are not unique. This, of course, does not
affect the  band structure which is basis independent, but becomes
an important  in the context of approximate many body treatments
(such as DMFT)  which include  only local Coulomb on the $f$
orbitals. For this purposes it is clearly advantageous to set up a
Hamiltonian in an orthogonal basis where the $f$ electrons are
highly localized. Hence, the secondary objective of this work is
to determine a good basis set for setting up models of the
actinides. For earlier tight-binding parametrization for actinides
see Ref.~\cite{Harrison83,Harrison87,Zhu2007}.

\subsection{Motivation for this work}

The idea of a localization-delocalization transition in the
actinides was brought forward by Johansson\cite{Johansson74}.
Johansson based this idea on an empirical comparison of the
canonical $5f$-bandwidth with the estimates of the Coulomb
interaction in the form of a Hubbard $U$, and therefore he is
starting with the assumption of a Hubbard model to represent the
actinides. As a result, the localization-delocalization transition
is designated as a Mott transition in his paper, but it should be
realized that this is a consequence of starting with an assumption
of a Hubbard model. For  a later elaboration of  these ideas in
the context of the $\alpha$-$\delta$ transition in Pu see
Ref.\cite{Katsnelson92,Severin92}.

The  important role of $d-f$ hybridization in actinide metals and
alloys was stressed in  the early work of Jullien et
al.\cite{Jullien72,Jullien73} who considered models similar to
Eq.~(\ref{model_Hamiltonian}).

In this paper, we reconsider the issue of the description of the
localization-delocalization transition in the actinide series from
a perspective which is motivated by recent DMFT and LDA+DMFT
works\cite{Shim2006,Pourovskii2007,Zhu2007}. These works utilize
the actinide hamiltonian (i.e. both $f$ and $spd$ states are
included), and have provided further demonstration of the
hypothesis that a localization-delocalization transition takes
place across the actinide series.  However, the relative
importance of $t^f$ and $V$ (i.e.  Hubbard-like vs. Anderson-like
contributions) were not explicitly examined in these studies.

\section{Orbitals and Basis}

\subsection{Basis set dependence issue}

While the issue of representing the Kohn-Sham hamiltonian in
different basis sets has been a subject of numerous studies, the
dependence of the results of  correlated electronic structure
methods such as LDA+DMFT on the choice of correlated orbitals is
only beginning to be explored\cite{solovyev}.

In this study, we investigate the role of the choice of the
correlated $f$ orbital. We first take the $f$ electron orbital as
the $f$ element of  the LMTO basis, both in the bare and screened
representations\cite{Skriver84,Andersen84}. The LMTO basis is
non-orthogonal and therefore must be orthogonalized in order to
avoid the complications of solving the many-body problem in a
non-orthogonal basis. As we will show in this study, the method of
orthogonalization has a large influence on the results. We utilize
both the L\"owdin orthogonalization\cite{Lowdin50} and the
projective orthogonalization\cite{Kristjan2006,Shim2006} that was
used in  earlier implementations of LDA+DMFT. This effectively
results in four different  constructions of $f$ orbitals, listed
in Table~\ref{four basis}.

\begin{table}
  \centering
  \renewcommand{\arraystretch}{1.3}
  \caption{Choice of basis.}
  \label{four basis}
  \begin{tabular}{| c | c |}
    \hline
    \,\,\, \textbf{ Bare LMTO} \,\,\, & \,\textbf{Screened LMTO}  \\
    \,\,\, L\"owdin transform & L\"owdin transform \\
    \hline
    \textbf{Bare LMTO} & \textbf{Screened LMTO}  \\
    projective basis & projective basis \\
    \hline
  \end{tabular}
\end{table}

\subsection{Bare and Screened LMTO within ASA scheme}

The basis set of linear muffin-tin orbitals (LMTOs) has been
extensively used in electronic structure calculations
\cite{Skriver84,Turek96}. Within the atomic sphere approximation
(ASA), LMTO is a minimal and efficient basis set with one basis
function per site $I$ and quantum pair $L=(l,m)$.  Although the
LMTO method is physically transparent, the constructed basis is
non-orthogonal.

Below we sketch the derivation of the bare and screened LMTO basis
set within the ASA. The construction of the bare LMTOs $\chi_{I
L}(\r)$ starts with so called envelope function \cite{Turek96},
which is a decaying solution of the Laplace equation centered at
the site $I$: \beq \label{Bare_Laplace_solution} K_{L}(\r_{I}) =
K_l (r_{I})Y_L(\hat{\r}_{I}) = \left(\frac{w}{r_{I}}
\right)^{l+1}Y_L(\hat{\r}_{I}),\eeq here $\r_{I}=\r-\R_I$, unit
vector $\hat{\r}_I$ indicates the direction of $\r_I$,
$Y_L(\hat{\r}_{I})$ is a spherical function, and $w$ is a scaling
parameter associated with the linear size of unit cell.

In any atomic sphere other than $I$, $K_{L}(\r_I)$ can be
represented as: \beq \label{tail_cancelation} K_{L}(\r_{I}) =
-\sum_{L^{\prime}} S_{I L, I^{\prime} L^{\prime}} J_{L^{\prime}}
(\r_{I^{\prime}}). \eeq The function $J_L(\r_{I})= (r_{I}/w)^l
Y_L(\hat{\r}_{I})$ stands for the regular solutions of Laplace
equation, and $S_{I L, I^{\prime} L^{\prime}}$ are structure
constants.

Inside each atomic sphere we construct a linear
combination of the solution $\phi_{I L}(\r_{I})$ of the Schr\"odinger
equation and its first derivative with respect to energy
$\dot{\phi}_{I L}(\r_{I})$ at some fixed energy $E_{\nu}$.

The final step is to smoothly match the boundary conditions at the
surface of sphere $I$: \beq \label{Bare_LMTO_BC_for_K}
\Phi^H_L(\r_I) \equiv A^K_{I L} \phi_{I L}(\r_{I}) + B^K_{I L}
\dot{\phi}_{I L}(\r_{I}) \rightarrow K_L (\r_{I})
 \eeq
and at the surface of sphere $I^{\prime}$ for all $I^{\prime} \neq
I$: \beq \label{Bare_LMTO_BC_for_J} \Phi^J_{L^{\prime}}
(\r_{I^{\prime}}) \equiv A^J_{I^{\prime} L^{\prime}}
\phi_{I^{\prime} L^{\prime}}(\r_{I^{\prime}}) +
B^J_{I^{\prime}L^{\prime}} \dot{\phi}_{I^{\prime}
L^{\prime}}(\r_{I^{\prime}}) \rightarrow J_{L^{\prime}}
(\r_{I^{\prime}}). \eeq

With the array of constants A and B determined from
(\ref{Bare_LMTO_BC_for_K}) - (\ref{Bare_LMTO_BC_for_J}) we
conclude the construction of bare LMTO basis function: \bea
\chi_{I L}(\r_{I}) = \left\{ \begin{array}{ll} \Phi^H_L(\r_I), &
\r_{I} \in S_{I}, \\   -\sum_{I^{\prime} L^{\prime}} S_{I L,
I^{\prime} L^{\prime}} \Phi^J_{L^{\prime}}
(\r_{I^{\prime}}), & \r_{I^{\prime}} \in S_{I^{\prime}}(I \neq I^{\prime}), \\
  K_L (\r_{I}), & \r \in Interstitial.
\end{array} \right. \\ \nonumber
\eea

The Fourier transform of the LMTOs with respect to $\R_I -
\R_{I^{\prime}}$ gives: \bea \label{k-space_LMTO_in_sphere}
\chi_{\k L}(\r) = \left\{ \begin{array}{ll} \Phi^H_L(\r) -
\sum_{L^{\prime}} \Phi^J_{L^{\prime}}(\r)S_{\k L L^{\prime}}, &  |\r| \le R_{MT}, \\
 \label{k-space_LMTO_out_sphere}  \sum_{\k} e^{i\k\R}
K_{L}(\r-\R), &  |\r| > R_{MT}. \\ \end{array} \right. \\
\nonumber \eea

The standard LMTO method outlined above yields long-range
orbitals. The concept of a screened LMTO was created to overcome
the non-locality of the bare LMTO basis set \cite{Andersen84}. The
method is based upon the idea of localizing the LMTOs by screening
with multipoles added on the neighboring spheres. Namely, to each
regular solution of Laplace equation we add  $-\alpha_{I L}$ of
the irregular solution: \beq \label{screening}
J^{\alpha}_{L}(\r_{I}) = J_{L}(\r_{I}) - \alpha_{I L}
K_{L}(\r_{I}).\eeq

The condition that on-site Laplace solution should not change
leads to the Dyson-like equation for the screened structure
constants: \beq \label{Dyson_equation_for_structure_const}
S^{\alpha}_{a, a^{\prime} } = S_{a,a^{\pp}}[\delta_{a^{\pp}
a^{\prime}}+\alpha_{a^{\pp}}S^{\alpha}_{a^{\pp}, a^{\prime}}
]=S_{a,a^{\pp}}U_{a^{\pp},a^{\prime}}, \eeq where matrix index $a$
refers to the pair $(I, L)$ and implies summation over repeated
indices. The matrices $\alpha_a \equiv \alpha_l$ are diagonal for
each $l$. In our calculations the choice of $\alpha$'s was as
follows: $\alpha_s=5.5166$, $\alpha_p=0.5242$, $\alpha_d=0.1382$
and $\alpha_f=0.0355.$

The screened and bare envelope functions are related by the
transformation $U_{a^{\prime},a}$ introduced in
(\ref{Dyson_equation_for_structure_const}): \beq
\label{Screened_Envelope_function} K^{\alpha}_L (\r_{I}) =
\sum_{I^{\prime} L^{\prime}} K_{L^{\prime}} (\r_{I^{\prime}})
[\delta_{ I^{\prime} L^{\prime}, I L}+\alpha_{I^{\prime}
L^{\prime}}S^{\alpha}_{ I^{\prime} L^{\prime}I L}], \eeq Or in
matrix notations: $$
K^{\alpha}_a=K_{a^{\prime}}U_{a^{\prime},a},$$ where $K_a\equiv
K_L(\r_I)$ and $K^{\alpha}_a\equiv K^{\alpha}_L(\r_I).$

 With the
definitions (\ref{screening}),
(\ref{Dyson_equation_for_structure_const}) and
(\ref{Screened_Envelope_function}) the construction of screened
LMTOs proceeds exactly in the same way as in the case of the bare
LMTOs. Namely, we construct new linear combinations $$\Phi^{H
\alpha}_L(\r_I) \equiv A^{K \alpha}_{I L} \phi_{I L}(\r_{I}) +
B^{K  \alpha}_{I L} \dot{\phi}_{I L}(\r_{I})$$ inside the sphere
$I$ by matching smoothly $K^{\alpha}_L (\r_{I})$ on its surface.
Also, we construct new linear combination $$\Phi^{J
\alpha}_{L^{\prime}}(\r_{I^{\prime}})\equiv A^{J
\alpha}_{I^{\prime} L^{\prime}} \phi_{I^{\prime}
L^{\prime}}(\r_{I^{\prime}}) + B^{J \alpha}_{I^{\prime}
L^{\prime}} \dot{\phi}_{I^{\prime} L^{\prime}}(\r_{I^{\prime}})$$
inside sphere $I^{\prime}$ by matching smoothly
$J^{\alpha}_{L^{\prime} }(\r_{I^{\prime}})$ on its surface for
each $I^{\prime} \neq I$. Thus, we arrive to the definition of the
screened LMTO: \bea \chi^{\alpha}_{I L}(\r_{I}) = \left\{
\begin{array}{ll} \Phi^{H \alpha}_L(\r_I), & \r_{I} \in S_{I}, \\
-\sum_{I^{\prime}L^{\prime}} S^{\alpha}_{I L, I^{\prime}
L^{\prime}} \Phi^{J \alpha}_{L^{\prime}}
(\r_{I^{\prime}}), & \r_{I^{\prime}} \in S_{I^{\prime}}(I \neq I^{\prime}), \\
  K^{\alpha}_L (\r_{I}), & \r \in Interstitial.
\end{array} \right. \\ \nonumber
\eea

The Fourier transform of the screened LMTOs with respect to $\R_I -
\R_{I^{\prime}}$ gives: \bea \label{k-space
screened_LMTO_in_sphere} \chi^{\alpha}_{\k L}(\r) = \left\{
\begin{array}{ll} \Phi^{H \alpha}_L(\r) -
\sum_{L^{\prime}} \Phi^{J \alpha}_{L^{\prime}}(\r)S_{\k L L^{\prime}}, &  |\r| \le R_{MT}, \\
 \label{k-space screened_LMTO_out_sphere}  \sum_{\k} e^{i\k\R}
K^{\alpha}_{L}(\r-\R), &  |\r| > R_{MT}. \\ \end{array} \right. \\
\nonumber \eea

The Hamiltonian and overlap matrices in screened and bare LMTO
representations  ($O$ ,  $H$, and $ O^{\alpha} $,  $H^{\alpha}$
respectively) are related through the transformation $U$
introduced in (\ref{Dyson_equation_for_structure_const}): \bea
H^{\alpha} = U^{\dagger} H U \\  O^{\alpha} = U^{\dagger} O U.
\label{def_overlap} \eea

Having constructed the basis, one has to solve the generalized eigenvalue problem:
\beq
\label{generalized_eigenvalue_problem}(H(\k)-\epsilon_i(\k)O(\k))\psi_i(\k)=0.
\eeq
As described above, it is necessary to transform to an orthogonal basis when performing
many-body calculations, such as DMFT, in order to avoid the difficulties associated with a
non-orthogonal basis.

\subsection{L\"owdin orthogonalization}

L\"owdin orthogonalization \cite{Lowdin50} is a
straightforward orthogonalization of the Hamiltonian which uses
no information from the basis set:
\beq \label{Lowdin transform}
\tilde{\mathcal{H}}(\k)=\frac1{\sqrt{O^{\dagger}(\k)}}
\mathcal{H}(\k) \frac1{\sqrt{O(\k)}}. \eeq

As will be shown below, this orthogonalization procedure may lead to
a further mixing of $L$ characters among the LMTOs and hence unphysical results.

\subsection{Projective orthogonalization}

A physically motivated orthogonalization  procedure is to find a
basis where each function contains the maximum amount of a
particular $L$ character. This  approach  proposed by  K. Haule
was  used in earlier LDA+DMFT studies of Cerium and Plutonium
\cite{Kristjan2006}. This basis has an important advantage, being
that  the "$f$ electron" in this basis has maximal $f$ character.
Mathematically, the non-interacting spectral function of the  $f$
electron Green's function in this basis agrees with the LDA
projected density-of-states having $f$ character, as shown in
Appendix A. This allowed us to identify the $f$ occupation in this
basis set with the occupation numbers inferred from EELS and X-Ray
absorption which are sensitive to angular momentum selection
rules\cite{moore}.

Here we follow Ref.\cite{Kristjan2006}. It is straightforward
using (\ref{k-space_LMTO_in_sphere}) and (\ref{def_overlap}) to
show that overlapping matrix within MT-sphere can be represented
as \cite{Andersen85}: \bea O_{\k L_1 L_2} = \delta_{L_1 L_2}
o^{(HH)}_{l_1} - S^{\dagger}_{\k
L_1 L_2} o^{(JH)}_{l_2} - o^{(HJ)}_{l_1} S_{\k L_1 L_2} \\
\nonumber + S^{\dagger}_{\k L_1 L^{\prime}} o^{(JJ)}_{l^{\prime}}
S_{\k L^{\prime} L_2}. \eea

The quantities $o^{HH}_l$, $o^{JH}_l$, $o^{HJ}_l$ and $o^{JJ}_l$
are numbers in each $l$-subspace. For $A$ and $B$ representing $H$
or $J$: \beq \label{def_small_o} o^{AB}_{l_1} = \la \Phi^A_{L_1}|
\Phi^B_{L_2} \ra \delta_{L_1L_2},\eeq

In each $L$-subspace the overlapping matrix is:
 \beq O_{\k } =
o^{(HH)} - S^{\dagger}_{\k } o^{(JH)} - o^{(HJ)} S_{\k } +
S^{\dagger}_{\k} o^{(JJ)} S_{\k }. \eeq In order to find the
transformation to the orthonormal base we must represent $O(\k)$ as
the square of a matrix. As we show below, the most intelligent choice
would be: \beq O(\k)\approx
(\mathcal{H}-\mathcal{J}S_{\k})^{\dagger}(\mathcal{H}-\mathcal{J}S_{\k})\eeq
for each $L$-subspace. Here $\mathcal{H}$ and $\mathcal{J}$ are
diagonal matrices proportional to unity in each subspace of
definite $L$ just like the overlaps $o^{(HH)}$ defined above.

The above equation can not be made exact  because the overlap numbers
are obtained by integration over the radial part of wave
functions. However, in most cases the overlap numbers can become
very close to their approximations: \beq
\begin{array}{ccc}
 \label{approximation}
o^{(HH)}_l & \approx & \mathcal{H}^*_l\mathcal{H}_l, \\
o^{(JH)}_l & \approx & \mathcal{J}^*_l\mathcal{H}_l, \\
o^{(HJ)}_l & \approx & \mathcal{H}^*_l\mathcal{J}_l, \\
o^{(JJ)}_l & \approx & \mathcal{J}^*_l\mathcal{J}_l. \\
\end{array}
\eeq

For each $L$ we have three independent equations for two unknowns.
An approximate solution can be found by minimizing the
following function: \bea |o^{(HH)}_l
-\mathcal{H}^*_l\mathcal{H}_l|^2 + |o^{(JH)}_l
-\mathcal{J}^*_l\mathcal{H}_l|^2 \\ \nonumber + |o^{(HJ)}_l
-\mathcal{H}^*_l\mathcal{J}_l|^2 + |o^{(JJ)}_l
-\mathcal{J}^*_l\mathcal{J}_l|^2 = min. \eea

The desired transformation to the new base is: \beq
T_{\k}=(\mathcal{H}- \mathcal{J}S_{\k})^{-1}.\eeq

Finally: \beq
\begin{array}{cccc}
\label{transformation}
O_{\k}^{new} & = & T^{\dagger}_{\k}O_{\k}T_{\k} & \approx 1, \\
H_{\k}^{new} & = & T^{\dagger}_{\k}H_{\k}T_{\k}.  \\
\end{array} \eeq

\section{Results}

We perform relativistic, spin-restricted LDA calculations within
the ASA scheme. $7s$, $6p$, $6d$ and $5f$-orbitals were chosen to
represent valence states, and 10$^3 \k$-points were used in the
first Brillouin zone. The same type of calculations were carried
out for 4 different materials, picked to evenly represent actinide
series: U, $\alpha$-Pu, $\delta$-Pu and Cm II ($fcc$ phase of
curium). For simplicity, we used the $fcc$ crystal structure for
each element. The lattice parameters listed in Table~\ref{lattice
parameters} were chosen to match the experimentally measured
volumes for corresponding phases in case of Pu and Cm II. For U we
use the equilibrium volume predicted within GGA calculations.

\begin{table}[]
  \centering
  \caption{Lattice parameters(in $\AA$).}
  \label{lattice parameters}
  \begin{tabular}{| l | l |}
    \hline
    \,\, $\alpha$-U & \,\, 4.3378 \,\,  \\
    \hline
    \,\, $\alpha$-Pu & \,\, 4.3074 \,\, \\
    \hline
    \,\, $\delta$-Pu & \,\, 4.6400 \,\, \\
    \hline
    \,\, Cm II & \,\, 4.9726 \,\, \\
    \hline
  \end{tabular}
\end{table}

\begin{figure}[]
\vspace{0.0cm}
\includegraphics[width=240pt,angle=0]{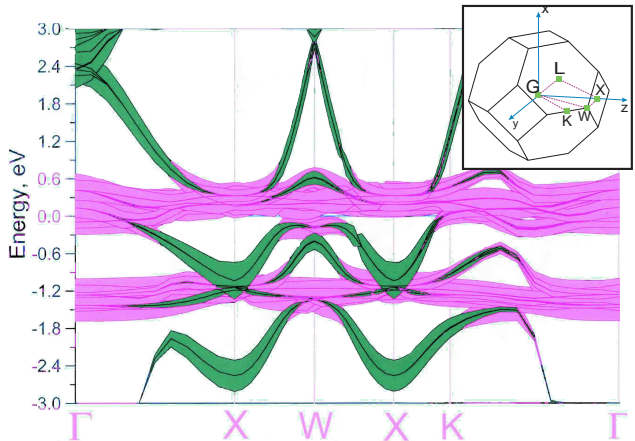}
\caption{(Color online) Band structure of Cm with indicated
contribution of $5f$ (light shade, red online) and $6d$ (dark
shade, green online) characters. The layout is chosen to show
$d$-contributions over $f$-contribution. Inset: Brillouin zone of
$fcc$ structure with indicated high symmetry directions. }
\label{Fat bands_Cm}
\end{figure}

In Figure~\ref{Fat bands_Cm} we present LDA band structure of Cm
II with the projections of the $5f$- and $6d$- characters. The
overwhelming contribution of $f$ character within a $1eV$ window around
Fermi level suggests the conclusion that the low-energy physics of
actinides is completely controlled by $f-f$ bonding. As we show
below this intuitive interpretation turns out to be misleading and
Hubbard model alone can not be considered as the low-energy
Hamiltonian for actinides. One has to account for presence of
$spd$-characters at the Fermi level through the hybridization.
Moreover, the hybridization energy scale in actinides turns out to
be larger than the average $f-f$ hopping.

\subsection{Determining a robust basis for the actinides}

\begin{widetext}
\begin{figure*}[t]
\vspace{0.0cm}
\includegraphics[width=360pt,angle=270]{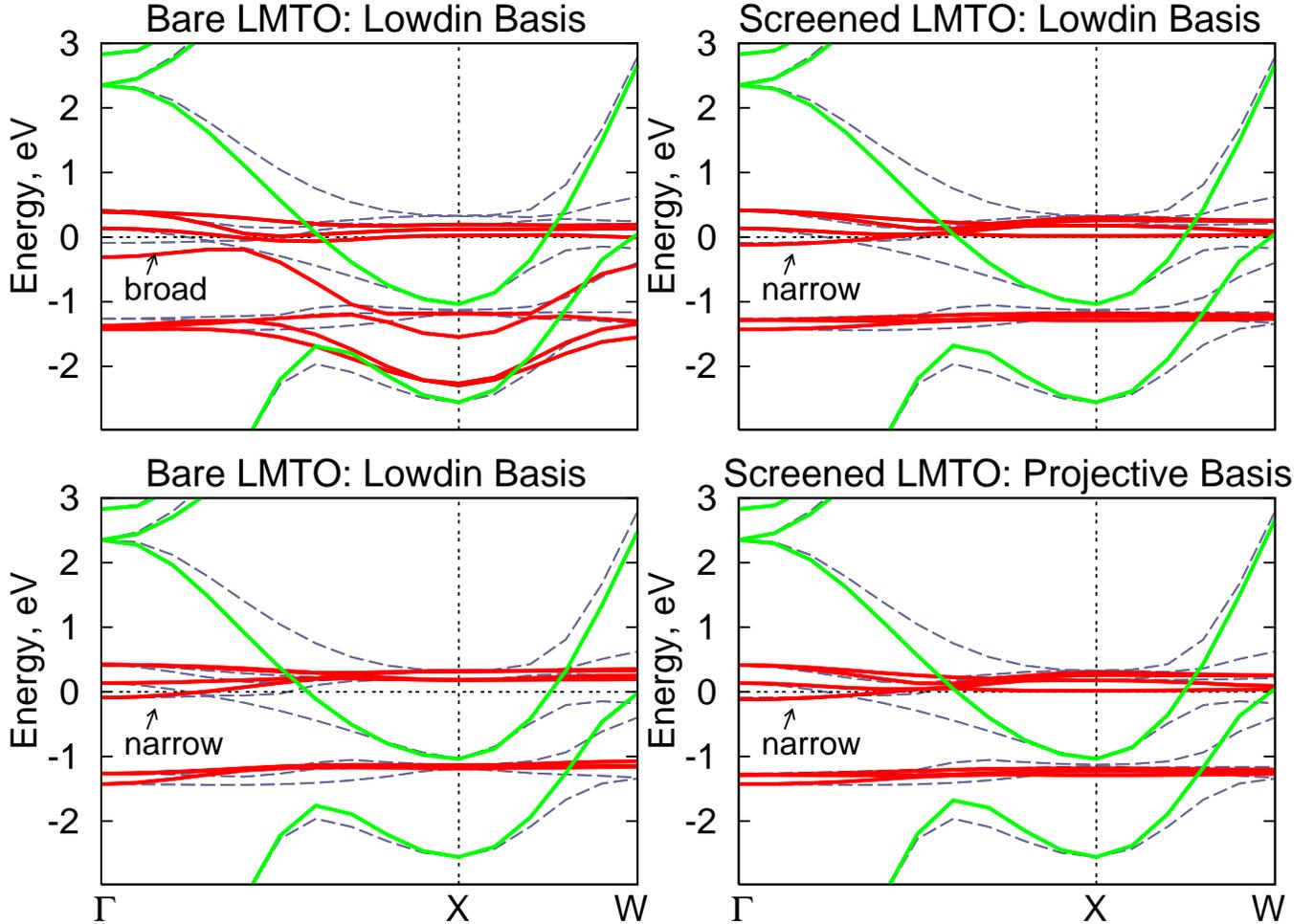}
\caption{(Color online) Basis difference for fcc Curium. In all
panels dashed grey lines represent the LDA band structure , solid
light shade lines (green online) represent bands for the block
$H_{spd}$ , and solid dark shade line (red online) represent bands
for the block  $H_f$. } \label{Basis difference}
\end{figure*}
\end{widetext}

In order to determine the optimum basis, we need to define a
criteria to judge the different bases. When performing DMFT
calculations, one accounts only for a subset of local electronic
correlations (those on the $f$ orbital). Therefore, from the
perspective of DMFT it is best to have $f$ orbitals with the
largest on-site Coulomb repulsion $U$ \cite{indranil}. A simpler
criteria, in the same spirit, is to search for  the smallest value
of $t^f$. We first investigate $t^f$ in Cm for the four different
basis sets in Table~\ref{four basis}. The hybridization $V$ in the
Hamiltonian (\ref{model_Hamiltonian}) may be set to zero. What
remains are the two blocks $H_f$ and $H_{spd}$ which are now
completely decoupled. The Hamiltonian may now be diagonalized
resulting in distinct $spd$ and $f$ bands, and any dispersion of
the $f$ bands is due to $t^f$. We begin by analyzing the bare
LMTOs orthogonalized with the L\"owdin procedure (see top left
panel Figure~\ref{Basis difference}). Some $f$ bands have a
dispersion greater than $1.5eV$ which is unfavorable. Using the
bare LMTOs orthogonalized with the projective procedure, the $f$
bands are far more narrow with a width of less than $0.4eV$ (see
left bottom panel of Figure~\ref{Basis difference}). In this case
the two sets of bands can be identified as $S=\frac{7}{2}$ and
$S=\frac{5}{2}$. The L\"owdin orthogonalization mixes the $spd$
states into the $f$ states which causes a larger dispersion and a
mixing of $f$ bands between the $S=\frac{7}{2}$ and
$S=\frac{5}{2}$ states. Alternatively, the projective
orthogonalization minimizes the amount of $spd$ character in the
$f$ states which results in weakly dispersing $f$ states.

The same exercise can be performed using the screened LMTOs (see
right top and bottom panels of Figure~\ref{Basis difference}). In
this case, both the L\"owdin and the projective orthogonalization
produce nearly identical results to the projective
orthogonalization of the bare LMTOs. The screened \hbox{LMTOs} are
insensitive to the method of orthogonalization due to the fact
that orbitals are already well localized with a well-defined
character.  In conclusion, one may use bare LMTOs orthogonalized
with the projective procedure or screened LMTOs orthogonalized in
an arbitrary manner as a robust basis for the actinides.

\subsection{Decomposition of the actinide band structures}

\begin{figure}[t]
\vspace{0.0cm}
\includegraphics[width=220pt,angle=0]{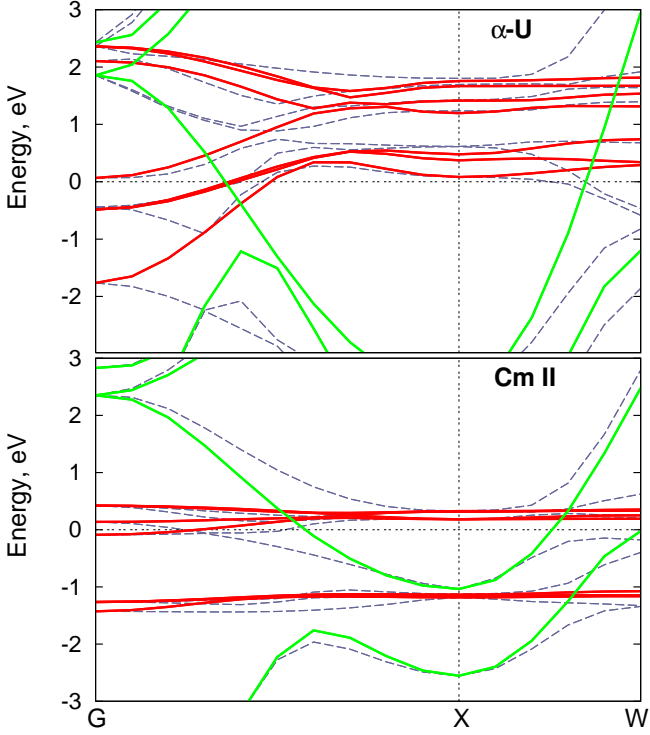}
\caption{(Color online) Band structure of $\alpha$-U (top) and Cm
II (bottom). Grey dashed lines represent LDA bands, solid light
shade lines (green online) represent bands of $H_{spd}$, and solid
dark shade lines (red online) represent bands of $H_f$.}
\label{Bnds_U_Cm}
\end{figure}

\begin{figure}[t]
\vspace{0.0cm}
\includegraphics[width=220pt,angle=0]{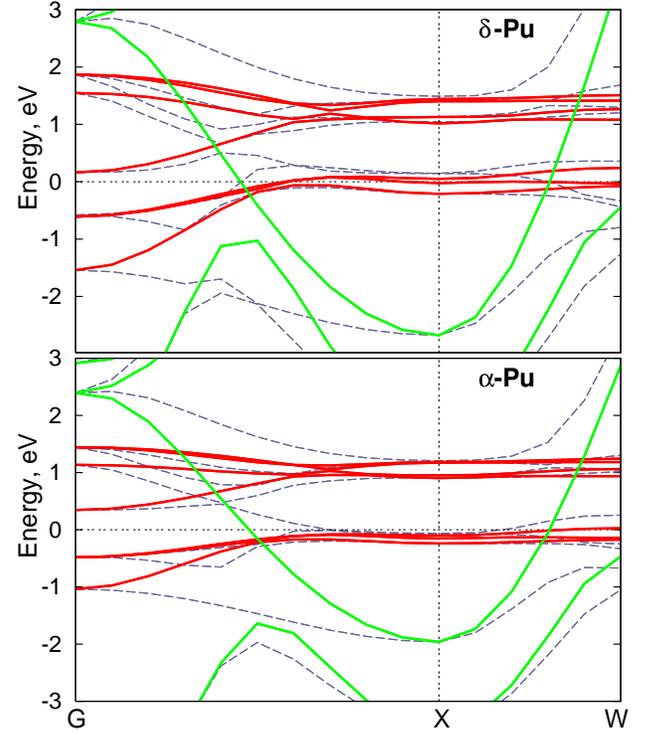}
\caption{(Color online) Band structure of $\alpha$-Pu (top) and
$\delta$-Pu(bottom). Grey dashed lines represent LDA bands,  solid
light shade lines (green online) represent bands of $H_{spd}$, and
solid dark shade lines (red online) represent bands of $H_f$. }
\label{Bnds_a_d_Pu}
\end{figure}

Having established a sensible basis for the actinides, we choose to
proceed with projective orthogonalization of bare LMTOs. It is
instructive to zero the hybridization $V$ of the Hamiltonian for
U, $\alpha$-Pu, $\delta$-Pu, and Cm, and to compare the full band
structure with the $spd$ and $f$ bands (see Figures
\ref{Bnds_U_Cm} and \ref{Bnds_a_d_Pu}).  The same generic behavior
can be seen in all four systems. The $spd$ bands have a strong
dispersion and cross the Fermi energy in all cases, and the $f$
bands are relatively narrow.  The fact that the $spd$ bands cross
the Fermi energy in all cases is a critical point which indicates
that there will be $spd$ states at the Fermi energy even if the
$f$ states become completely localized.  When the hybridization
$V$ is switched on, the $f$ and $spd$ bands interact via $V$ and
mix. Therefore the strength of $V$ can qualitatively be seen as
the difference between the full DFT bands and the $f$+$spd$ bands.
The $f$ bands are relatively wide
for Uranium and become increasingly narrow as Curium is approached.
The $spd$ bands follow the same general trend, but the relative changes are smaller.
The values of $V$ and $t^f$ will be
quantified below.

\subsection{Quantitative analysis of $V$ and $t^f$}
In order to quantify $V$ and $t^f$ for the different actinides, we
introduce an average $V$ and $t^f$ so each actinide may be
characterized by two numbers.

First, we recall that the Hamiltonian (\ref{model_Hamiltonian})
consist of four blocks:\bea \label{blocks of Hamiltonian} H(\k)
=\left(
\begin{array}{cc} H^{spd}(\k) & V_{\k} \\ V_{\k}^{\dagger} & H^{f}(\k) \\
\end{array} \right)\eea

Then the average strength of the hybridization per band is defined
as follows: \beq \label{Average Hybridization}
\overline{V}=\frac1{N_f}[\frac12 \emph{Tr}\la
\tilde{H}(\k)\tilde{H}(\k) \ra]^{1/2}, \eeq where $\tilde{H}(\k)$
stands for hamiltonian (\ref{blocks of Hamiltonian}) with
$H_{spd}(\k)=H_{f}(\k)=0$, $N_f=14$ stands for number of
$f$-bands, and $\la \ldots \ra = \frac1{N_{\k}}\sum_{\k}\ldots$ .
The definition (\ref{Average Hybridization}) was chosen to match
hybridization $V$ of standard Anderson model in two-band limit.

The average value of  $t^{f}$  is defined as follows:
    \beq \overline{t^{f}}=\frac1{N_f}[\emph{Tr} (\la H_f(\k)^2\ra  - \la H_f(\k)\ra ^2)]^{1/2},\\
    \label{Average t} \eeq
and matches $t^f$ of the canonical Hubbard model in the limit of
one-band model.

Table ~\ref{ff hoppings vs hybridization} lists the calculated values
of the average hybridization $\overline{V} $ and
$\overline{t^{f}}$ and the average energy for the $j=5/2$ and
$j=7/2$ levels of the $f$ manifold relative to the Fermi energy. The
averages are generally the same for the  bare and screened
LMTOs, with the exception of the average hybridization being
slightly larger in the case of screened LMTOs.

\begin{center}
\begin{table}
  \centering
  \caption{Quantitative characteristics for actinide series (in eV).}
  \label{ff hoppings vs hybridization}
  \begin{tabular}{| l | l | l | l | l | l |}
    \hline
    & \,\,\, $\overline{V}$\,\,\, &\,\,\,$\overline{t^{f}}$ &  \,\,\,$\overline{V} / \overline{t^{f}}$ & \,\,\,$\epsilon_{5/2} -\mu $ & \,\,\,$\epsilon_{7/2} -\mu$ \\
    \hline
    \multicolumn{6}{|c|}{\bf Bare LMTO}\\
    \hline
    \,\,\, $\alpha$-U & \,\,\, 0.483 \,\,\, & \,\,\, 0.188 \,\,\, &  \,\,\, 2.569 \,\,\, &  \,\,\, 0.442 \,\,\, &  \,\,\,1.353  \,\,\, \\
    \hline
    \,\,\, $\alpha$-Pu & \,\,\, 0.423 \,\,\, & \,\,\, 0.146 \,\,\, &  \,\,\, 2.897   \,\,\, &  \,\,\, -0.180 \,\,\, &  \,\,\, 0.971 \,\,\, \\
    \hline
    \,\,\, $\delta$-Pu & \,\,\, 0.305 \,\,\, & \,\,\, 0.099 \,\,\, &  \,\,\, 3.081 \,\,\,  &  \,\,\, -0.129 \,\,\, &  \,\,\, 1.008 \,\,\, \\
    \hline
    \,\,\, Cm II & \,\,\, 0.189 \,\,\, & \,\,\, 0.050 \,\,\, &  \,\,\, 3.780 \,\,\,  &  \,\,\,-1.152 \,\,\, &  \,\,\, 0.238 \,\,\,  \\
    \hline
    \hline
    \multicolumn{6}{|c|}{\bf Screened LMTO}\\
    \hline
    \,\,\, $\alpha$-U & \,\,\, 0.490 \,\,\, & \,\,\, 0.188 \,\,\, & \,\,\, 2.606 \,\,\, &  \,\,\, 0.444 \,\,\, &  \,\,\, 1.355 \,\,\,  \\
    \hline
    \,\,\, $\alpha$-Pu & \,\,\, 0.429 \,\,\, & \,\,\, 0.146 \,\,\, & \,\,\, 2.938  \,\,\, &  \,\,\,  -0.178 \,\,\, &  \,\,\,  0.973 \,\,\, \\
    \hline
    \,\,\, $\delta$-Pu & \,\,\, 0.309 \,\,\, & \,\,\, 0.098 \,\,\, & \,\,\, 3.153  \,\,\, &  \,\,\, -0.128 \,\,\, &  \,\,\, 1.009 \,\,\,  \\
    \hline
    \,\,\, Cm II & \,\,\, 0.192 \,\,\, & \,\,\, 0.050 \,\,\, & \,\,\, 3.840 \,\,\, &  \,\,\, -1.151  \,\,\, &  \,\,\, 0.238 \,\,\,   \\
    \hline
  \end{tabular}
\end{table}
\end{center}

\begin{center}
\begin{table}
  \centering
  \caption{Nearest Neighbors contributions to $\overline{V}$ and $\overline{t^{f}}$ (in eV).}
  \label{nearest_neighbors}
  \begin{tabular}{| l | l | l | l |}
    \hline
    & \,\,\, $\overline{V}_{n}$\,\,\, &\,\,\,$\overline{t^{f}_{n}}$ &  \,\,\,$\overline{V_{n}}/ \overline{t^{f}_{n}}$  \\
    \hline
    \,\,\, $\alpha$-U & \,\,\, 0.371 \,\,\, & \,\,\, 0.172  \,\,\, & \,\,\, 2.157  \,\,\, \\
    \hline
    \,\,\, $\alpha$-Pu & \,\,\, 0.324 \,\,\, & \,\,\, 0.134 \,\,\, & \,\,\, 2.418  \,\,\,  \\
    \hline
    \,\,\, $\delta$-Pu & \,\,\, 0.232 \,\,\, & \,\,\, 0.090 \,\,\, & \,\,\, 2.578  \,\,\,   \\
    \hline
    \,\,\, Cm II & \,\,\, 0.143 \,\,\, & \,\,\, 0.045 \,\,\, & \,\,\, 3.178 \,\,\,   \\
    \hline
  \end{tabular}
\end{table}
\end{center}

\begin{figure}[t]
\vspace{1.0cm}
\includegraphics[width=160pt,angle=270]{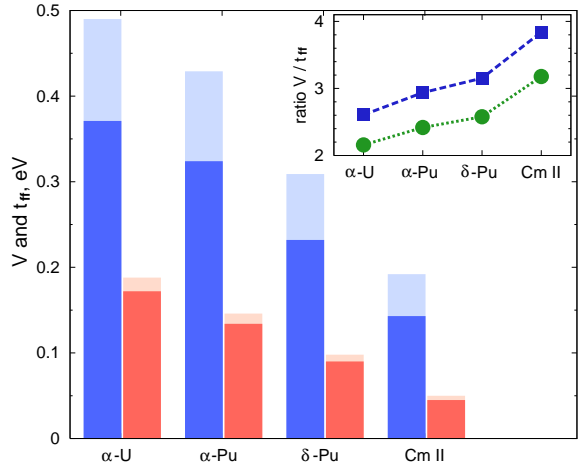}
\caption{(Color online) Histogram represents average hybridization
(first bar for each element, blue online) and average $f-f$
hopping (second bar for each element, red online) as functions of
atomic number. The shadow bars show $\overline{V}$ and
$\overline{t^{f}}$ for the original Hamiltonian and bright bars
represent values $\overline{V}_{n}$ and $\overline{t^{f}_{n}}$
arising from nearest neighbors contributions only. In inset: the
ratio $\overline{V}/\overline{t^{f}}$ (squares) as function of
atomic number. The ratio of nearest neighbor contributions is
represented by circles. } \label{V_t_histogram}
\end{figure}

These results are displayed graphically in Figure \ref{V_t_histogram}.
In all cases, $\overline{V}$ is significantly greater than $\overline{t^f}$.
As one moves along actinides series from U to Cm
$\overline{t^f}$ decreases as much as four times. The average value of
hybridization $\overline{V}$ also decreases but at a slower rate, as
indicated by the inset plot of the ratio of $\overline{V}$ and $\overline{t^f}$.
The strong decrease in $\overline{V}$ and $\overline{t^f}$ will both contribute
to the localization of the $f$ states. In order to determine if the localization could
be predominantly assigned to either Mott or Anderson character, explicit many-body
calculations such as DMFT would need to be performed.

In order to provide further insight into the degree of locality of
the basis, it is instructive to determine the fraction of
$\overline{V_{n}}$ and $\overline{t^f_{n}}$ which arise solely
from nearest-neighbor hopping. The corresponding values are
presented in Table~\ref{nearest_neighbors} and can also be seen in
Figure \ref{V_t_histogram}. First nearest neighbors contribute
$\approx 75\%$ to $\overline{V}$ and $\approx 90 \%$ to the
$\overline{t^f}$. The ratio of
$\overline{V_{n}}/\overline{t^f_{n}}$ is also given for the
nearest-neighbor contribution, and the shape and slopes of the two
respective curves are very similar. This analysis indicates that
nearest-neighbor hopping in real space accounts for most of the
relevant one-electron physics.

\subsection{Real space analysis of band structure}

\begin{figure}[t]
\vspace{0.0cm}
\includegraphics[width=180pt,angle=270]{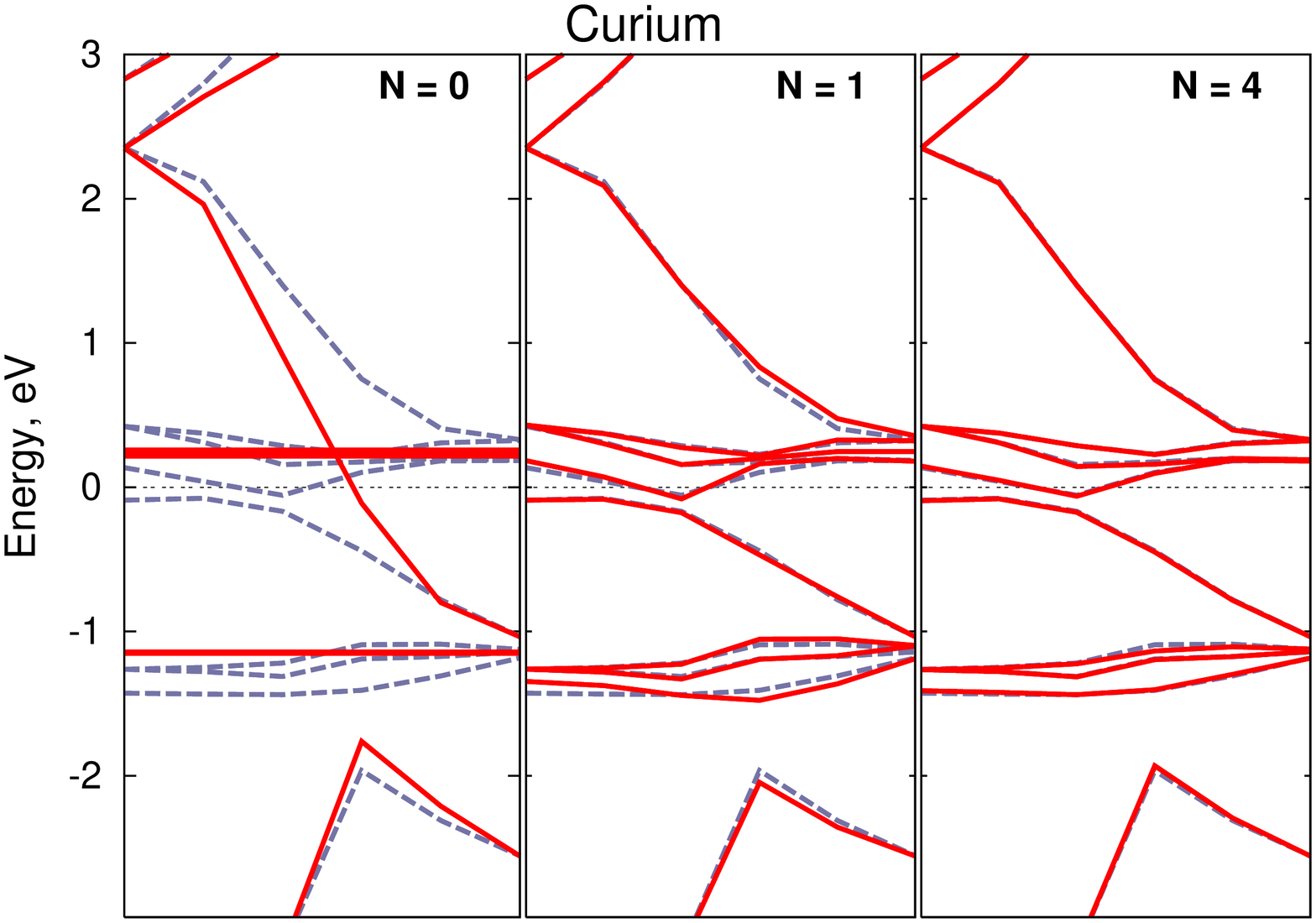}
\vspace{0.0cm}
\includegraphics[width=180pt,angle=270]{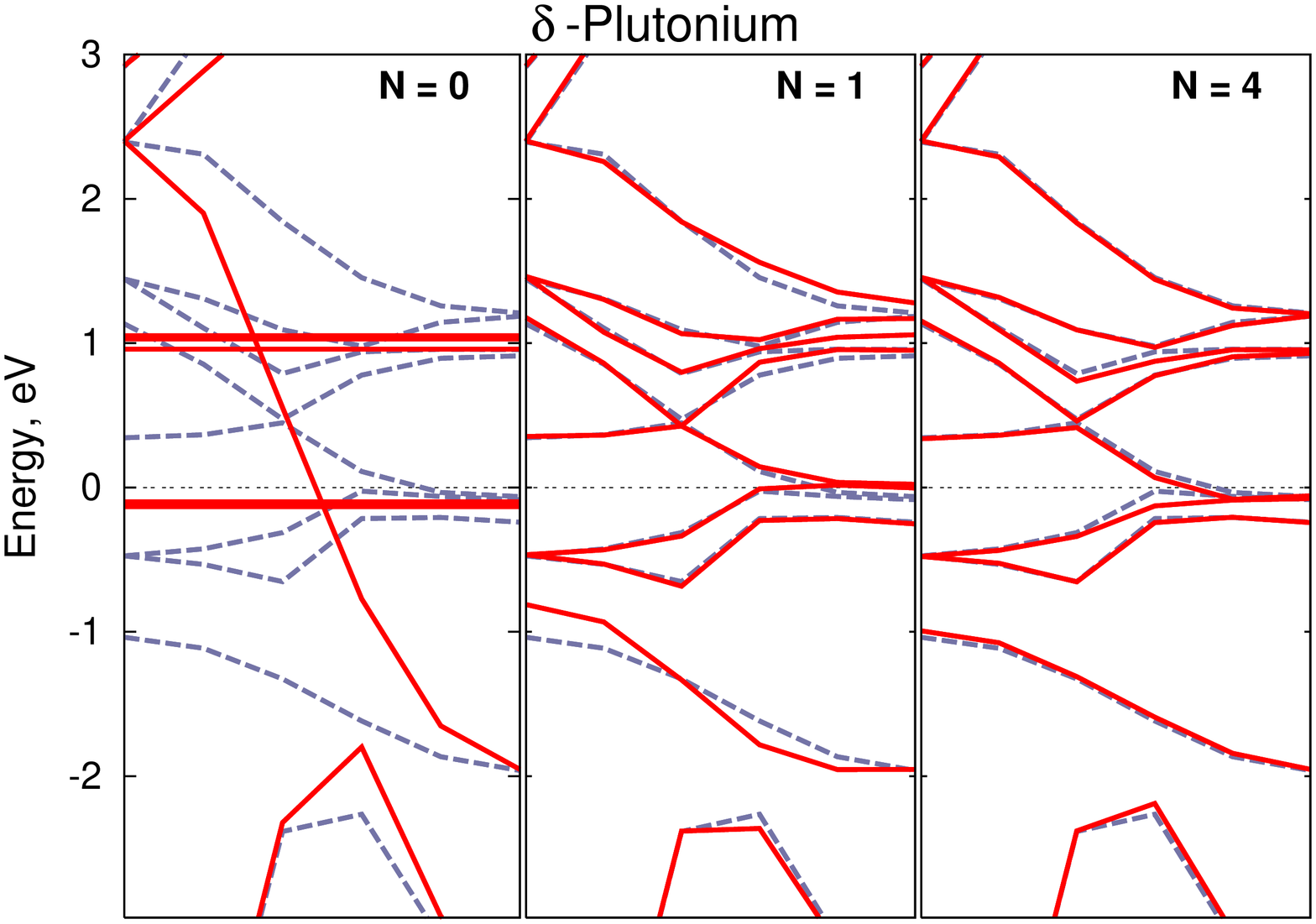}
\caption{(Color online) Band structure of Curium (top) and
$\delta$-Pu (bottom) when N nearest neighbors are taken into
account for $f$-orbitals (red solid line) is compared to original
LDA bands (dashed grey line). The band structures are plotted for
$\Gamma$-$X$ direction. } \label{Real_space_Pud} \vspace{0.0cm}
\end{figure}

\begin{figure}[t]
\vspace{0.5cm}
\includegraphics[width=180pt,angle=270]{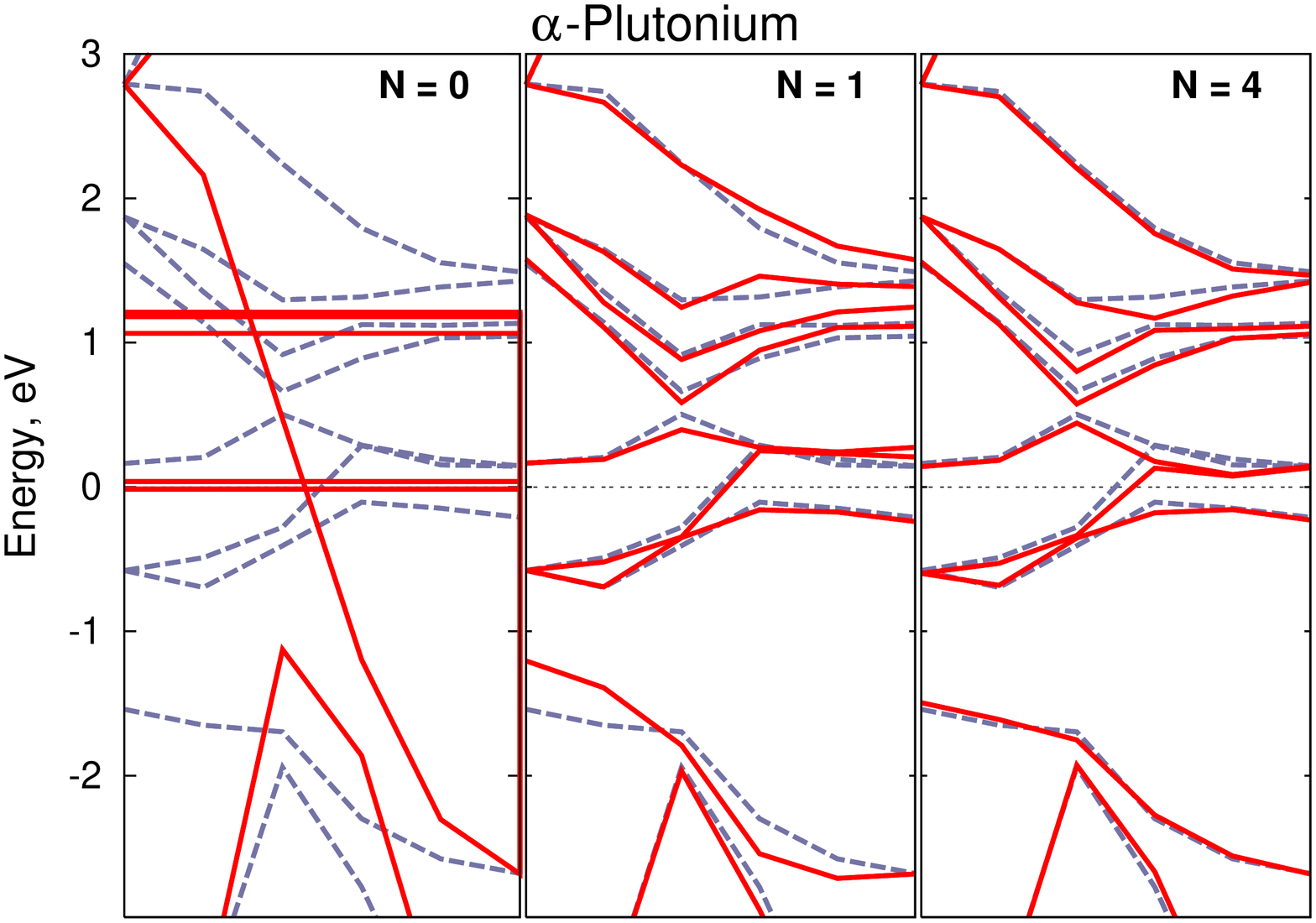}
\includegraphics[width=180pt,angle=270]{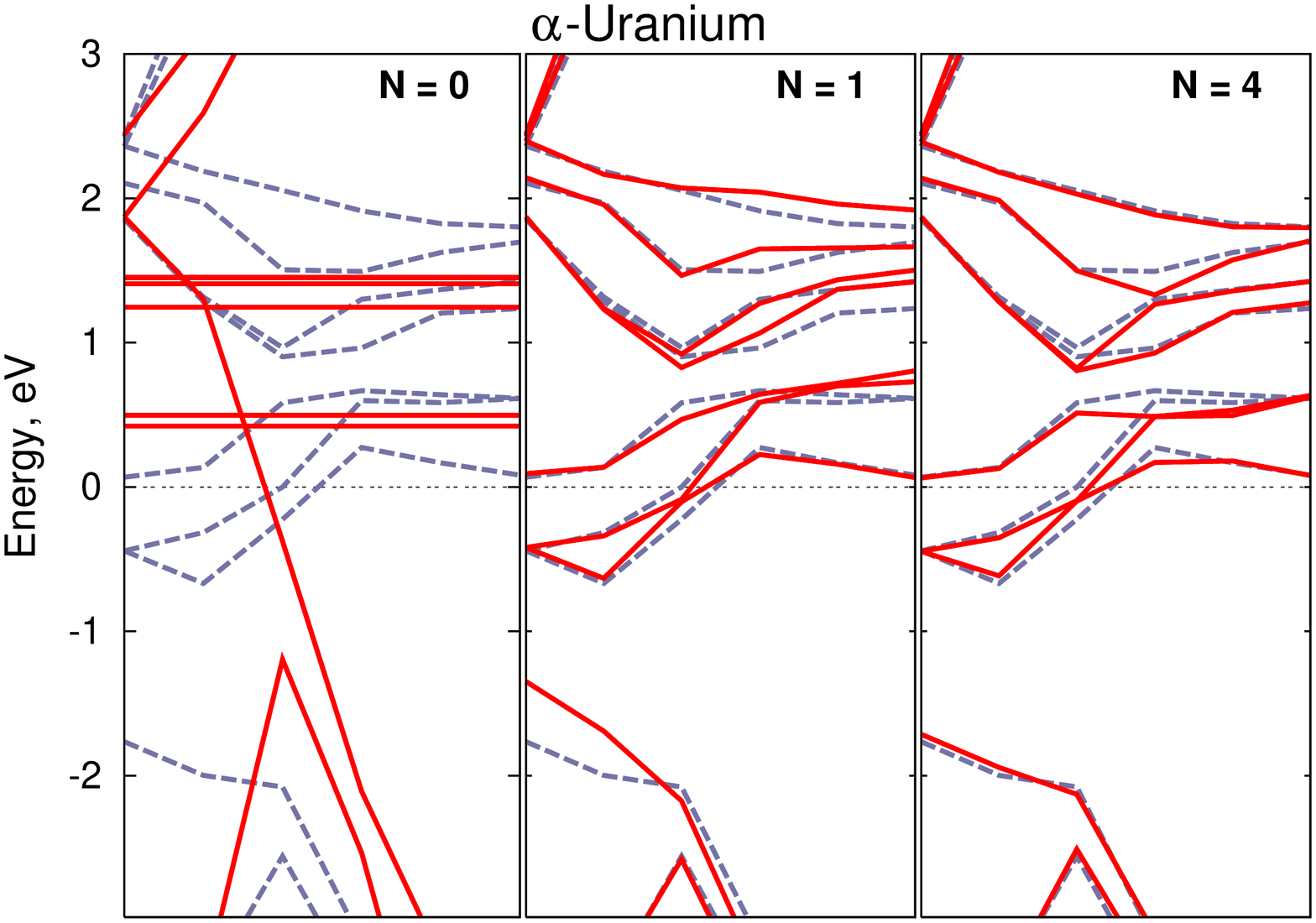}
\caption{(Color online) Band structure of $\alpha$-Pu (top) and
$\alpha$-U (bottom) when N nearest neighbors are taken into
account for $f$-orbitals (red solid line) is compared to original
LDA bands (dashed grey line). The band structures are plotted for
$\Gamma$-$X$ direction. } \label{Real_space_U}
\end{figure}

In the above analysis it was shown that nearest-neighbor hopping
accounts for a strong majority of $\overline{V}$ and
$\overline{t^{f}}$. Therefore, it is suggestive that the
one-electron bands can be reproduced with relatively short ranged
hoppings $t^f$ and $V$. In order to determine the degree of
locality, we plot the band structure as a function of the number
of neighbors for the $t^f$ and $V$ hoppings (see Figures
 \ref{Real_space_Pud} and \ref{Real_space_U}). Results are given for zero, one, and fourth
neighbor hopping. The hopping $t^{spd}$ are not truncated as it is
clear they will definitely have relatively long-range hoppings.
The generic results are similar for all four materials. The case
with zero neighbors results in flat bands having the on-site
energy of each orbital. When first nearest-neighbors are included
the resulting bands are an excellent approximation to the full
band structure. Including up to four nearest-neighbors yields
nearly perfect agreement.  Cm has better agreement than U for a
given number of nearest-neighbors, and this reflects the larger
degree of localization in the late actinides as compared to the
early actinides. In conclusion, the band structure of the
actinides is dominated by nearest-neighbor hopping when using an
appropriate basis.

\subsection{Conclusion}

In summary, a one-electron analysis of band structure of the
actinides was presented. We demonstrated that bare LMTOs
orthogonalized with the projective method and screened LMTOs are
robust bases, in the sense that they give rise to $f$ orbitals
with minimal hopping. Analysis of the Hamiltonian in these bases
yielded a number of interesting results. When switching off the
hybridization $V$, it was shown that the $spd$ states cross the
Fermi energy and hence will be present at the Fermi energy even if
the $f$ electrons become localized. Our description is in
reasonable agreement with the earlier work of
Harrison\cite{Harrison83}. In particular, the matrix elements of
spin-orbit coupling indeed have atomic-like nature and the
hybridization is much  larger than the direct $f-f$ hopping
\cite{Harrison83}. However  the $spd$ bands are not simple plane
waves and the hybridization matrix element does not have a simple
$\k$-dependence proportional to an $l=2$  spherical harmonic.

Evaluation of the average hybridization $\overline{V}$ and average
$f-f$ hopping $\overline{t^f}$ as a function of the actinides
showed that both quantities decrease strongly. The quantity
$\overline{t^f}$ decreased faster than $\overline{V}$, but
$\overline{V}$  was larger in all actinides. Hence, the Anderson
model  of the localization-delocalization transition, rather than
a multiorbital Hubbard model  is needed to describe the physics of
the actinides once explicit many-body calculations are added. This
is the point of view taken in recent DMFT work \cite{Shim2006},
and no further reduction to a model containing only f bands seems
possible. Finally, a real-space analysis of the band structure
demonstrated that nearest-neighbor hopping accounts for most of
the band structure in the basis used in this study, thus providing
a tight binding fitting of the bands of the actinides that can be
useful in further studies.

\acknowledgments
This research was sponsored by the National Nuclear Security Administration under the Stewardship Science Academic Alliances program
through DOE Research Grant  DE-FG52-06NA26210. We are grateful to S. Savrasov for useful
discussions.

\appendix
\section{Green's function in the projective base}

In the DMFT approach one needs to choose a set of localized Wannier
states in which correlations are strongest. In the context of
actinides, the orbitals with the largest component of $5f$ character
are the appropriate set of orbitals.

For many-body calculations, it is convenient if the set of
localized orbitals is orthogonal. In this case, it is desired that
the local Green's function is connected to the partial density of
states by the usual relation
\begin{equation}
D^l_{mm'}(\omega) \approx \frac{1}{2\pi i}(\widetilde{G}^\dagger_{loc}-\widetilde{G}_{loc})_{lm,lm'}.
\label{DOSG}
\end{equation}
In another words, the localized set of orbitals need to give rise to the
$5f$ spectra defined by
\begin{equation}
  D^l_{mm'}(\omega) = \int \frac{d\vr_1 d\vr_2}{2\pi i}
  Y_{lm}^*(\hat{\vr}_1)
  (G^\dagger(\vr_1,\vr_2)-G(\vr_1,\vr_2))Y_{lm'}(\hat{\vr}_2)
\end{equation}
with $l=3$ for actinides. Only in this case, the number of $f$
electrons (or the valence of the material) is connected to the
impurity $f$ count, as obtained in the DMFT calculation.

Using the LMTO basis set Eq.~(\ref{k-space_LMTO_in_sphere}), the
partial density of states becomes
\begin{widetext}
\begin{eqnarray}
D^l_{mm'}(\omega) = \frac{1}{2\pi i}\sum_{\k}
\left\{
(G_{\k\omega}^\dagger-G_{\k\omega}) o^{HH}
-[S_\k (G_{\k\omega}^\dagger-G_{\k\omega})] o^{HJ}
-[(G_{\k\omega}^\dagger-G_{\k\omega})S_\k^\dagger] o^{JH}
+[S_\k (G_{\k\omega}^\dagger-G_{\k\omega})S_\k^\dagger] o^{JJ}
\right\}_{lm,lm'}
\label{PDOS}
\end{eqnarray}
\end{widetext}
where the momentum dependent Green's function is
\begin{equation}
G_{\k\omega} = (O_\k(\omega+\mu)-H_\k-\Sigma_{\k\omega})^{-1}
\end{equation}
and overlap numbers $o^{AB}$ are defined by Eq.~(\ref{def_small_o}).

The projective orthogonalization Eq.~(\ref{transformation}) leads to the
following Green's function
\begin{eqnarray}
\widetilde{G}_{\k\omega} = (T_\k^\dagger [O_\k(\omega+\mu)-H_\k-\Sigma_{\k\omega}] T_\k)^{-1}\\
= (\mathcal{H}-\mathcal{J} S_\k)G_{\k\omega}(\mathcal{H}-\mathcal{J}S_\k)^\dagger.
\end{eqnarray}
The local spectral function in this new base therefore becomes
\begin{widetext}
\begin{eqnarray}
\frac{1}{2\pi i}(\widetilde{G}^\dagger_{loc}-\widetilde{G}_{loc})_{lm,lm'}&=&
  \frac{1}{2\pi i}\sum_\k\left\{ [G^\dagger_{\k\omega}-G_{\k\omega}]_{LL'}\mathcal{H}_l^*\mathcal{H}_l
  -[S_\k (G_{\k\omega}^\dagger-G_{\k\omega})]_{LL'}\mathcal{H}_l^*\mathcal{J}_l
  \right.\nonumber\\
  &&\left.
  \quad-[(G^\dagger_{\k\omega}-G_{\k\omega})S_\k^\dagger]_{LL'}\mathcal{J}_l^*\mathcal{H}_l
  +[S_\k (G_{\k\omega}^\dagger-G_{\k\omega})
  S_\k^\dagger]_{LL'}\mathcal{J}_l^*\mathcal{J}_l
  \right\}
\end{eqnarray}
\end{widetext}
which is equivalent to the partial density of state Eq.~(\ref{PDOS}) provided the condition
Eq.~(\ref{approximation}) is satisfied.
Extensive experience shows that in the case of
localized $d$ and $f$ orbitals, the condition is always
satisfied to very high accuracy (better than $1\%$) therefore the
relation between the partial density of states and local Green's
function Eq.~(\ref{DOSG}) is also satisfied to high accuracy.

\end{document}